# Systematic efficiency study of line-doubled zone plates

F. Marschall[a], J. Vila-Comamala[a], V. A. Guzenko[a], C. David[a]

[a] Paul Scherrer Institut, CH-5232 Villigen PSI, Switzerland

## Abstract

Line-doubled Fresnel zone plates provide nanoscale, high aspect ratio structures required for efficient high resolution imaging in the multi-keV x-ray range. For the fabrication of such optics a high aspect ratio HSQ resist template is produced by electron-beam lithography and then covered with Ir by atomic layer deposition (ALD).

The diffraction efficiency of a line-doubled zone plate depends on the width of the HSQ resist structures as well as on the thickness of the deposited Ir layer. It is very difficult to measure these dimensions by inspection in a scanning electron microscope (SEM) without performing laborious and destructive cross-sections by focus ion beams (FIB). On the other hand, a systematic measurement of the diffraction efficiencies using synchrotron radiation in order to optimize the fabrication parameters is not realistic, as access to synchrotron radiation is sparse.

We present a fast and reliable method to study the diffraction efficiency using filtered radiation from an X-ray tube with a copper anode, providing an effective spectrum centered around 8 keV. A large number of Fresnel zone plates with varying dimensions of the resist structures and the ALD coating were measured in an iterative manner. Our results show an excellent match with model calculations. Moreover, this systematic study enables us to identify the optimum fabrication parameters, resulting in a significant increase in diffraction efficiency compared to Fresnel zone plates fabricated earlier without having feedback from a systematic efficiency measurement.

## Introduction

Fresnel zone plates (FZPs) serve as diffractive lenses in a variety of X-ray microscopy techniques [1, 2, 3, 4]. They can either be used in scanning X-ray microscopes to produce a small, intense focal spot that is scanned across a sample, or as objective lenses in full-field X-ray microscopes to create a magnified x-ray image. FZPs consist of many concentric rings with pitches that decrease from the center to the outer edge. A coherent incident plane x-ray wave will be diffracted by this pattern in such a way, that the first order diffraction interferes constructively in a focal point. Apart from the fact that several diffraction orders exist, a FZP acts very similar to a normal lens. The achievable spatial resolution is limited to approximately the half pitch of the outermost zones. Therefore, FZPs have to be made with small lateral structure sizes, when high resolution is needed. Presently, FZP-based x-ray microscopes routinely achieve sub-100 nm resolution, and in some cases, the resolution is approaching the 10 nm level [5, 6, 7, 8].

Apart from providing sufficiently high resolution, FZPs used in the x-ray range should also give sufficient diffraction efficiency. The optimum efficiency is obtained when the zone plate structures induce a phase shift close to $\pi$ to the x-ray wave. For x-ray optics used at multi-keV photon energies, this means that zone structures should be made of heavy materials and be several 100 nm high. Consequently, nano-structures with very high aspect ratios need to be made by nanolithography techniques. These include the generation of high aspect ratio polymethyl-methacrylate (PMMA) resist structures, which are then filled by gold by electroplating. FZP structures with aspect ratios up to 15 have been achieved using this method [9, 10, 11].

A successful way produce metallic nano-structures with smaller line widths and higher aspect ratios is the so-called line-doubling technique [12, 13]. It is based on the fabrication of a template structures made of a light material having only a weak effect on x-rays, which is coated with a heavy metal by ALD , as shown in Figure 1. As the X-ray mainly interact with the metal deposited on the side walls of the template structures, the effective number of lines is doubled. In our case the template is written into HSQ (Hydrogen silsesquioxane) and then coated with Ir (see Figure 2).

High aspect ratio nano-structures can also be fabricated by other techniques like MACE (Metal Assisted Chemical Etching) in combination with line-doubling [14], multilayer lenses [15-18] or by stacking [19]. However, reaching the highest possible aspect ratio is not in the focus of this manuscript. Here we will focus on the effect of the duty cycles in line-doubled FZPs, which is independent from its actual line width and structure height.

## Efficiency calculations

In order to investigate the effect of the geometry of a diffractive x-ray optical element, we first consider the case of a diffraction grating with constant period. Its efficiency can be calculated using the x-ray optical constants tabulated by Henke et al. [21]. It mainly depends on the height of the structures and the duty cycle, which is defined as the ratio of the pitch to the line width. In this study we focus on the effect of the duty cycle. We consider zone plates consisting of 550 nm high HSQ template structures that are conformally coated with an Ir layer. This geometry gives maximum diffraction efficiency at a photon energy of 4.5 keV. However, in our study we consider a photon energy of 8 keV, which still results in appreciable diffraction efficiencies and allows for convenient characterization with radiation from an x-ray tube source. It should be noted that the actual height or photon energy has no impact on the optimum duty cycle.

For a binary grating, with alternating empty and filled lines, the efficiency is highest for a duty cycle of 0.5. In a line-doubled grating, the situation is more complex, as the structure is a periodic arrangement of zones, that each consist of four parts: air, Ir, HSQ, Ir. Taking into account that the two Ir parts are of equal width, such a grating has two duty cycles, one concerning the HSQ structure width and one concerning the Ir layer thickness.

Figure 3 shows the effect of both duty cycles on the diffraction efficiency, assuming perfectly shaped zones. The first order diffraction efficiency of a line-doubled grating is highest, when all parts have the same size, which means that both, Ir duty cycle and HSQ duty cycle, are 0.25. All deviations from this optimum lead to decrease of the first order diffraction efficiency and to increase of other orders. Especially the half order, which is just the first order of the HSQ template, can become significant if wrong fabrication parameters are chosen.

The situation is again somewhat different when considering the situation of a FZP, as the pitch of a FZP is changing over its radius. As Ir is deposited by ALD, its thickness is constant over the entire FZP. Thus, Ir duty cycle is changing with the radius. Consequently the efficiency of a line-doubled FZP is changing over its radius, as shown in Figure 4. Depositing a little more Ir as optimum for the outermost zones, shifts the maximum diffraction efficiency towards the center of the FZP. Thus the overall diffraction efficiency of a line-doubled FZP can be increased. Figure 5 shows the integrated diffraction efficiency of a line-doubled FZP dependent on both duty cycles. The result of these calculations is, that for a FZP the highest diffraction efficiency can be achieved by an HSQ duty cycle of 0.21 and an Ir duty cycle of 0.33.

## Fabrication of Fresnel zone plates (FZPs)

For the experiment we produced membrane chips each having 16 zone plates. All zone plates had the very same diameter of 150 µm, and an effective outermost half pitch of the Ir structures of 30 nm. This is not the smallest achievable structure size using the line-doubling technique, but was very suitable for this study, as the smallest achievable structure size was not goal of the optimization.

The FZPs for our experiment were fabricated on silicon nitride membranes. For the membrane fabrication, double side polished silicon wafers with 250 µm thickness were coated with 250 nm low stress silicon nitride on both sides. The silicon nitride on the rear side of the wafers was structured by photolithography and reactive ion etching. Using the structured silicon nitride as mask, the silicon was anisotropicly etched in 20 % KOH at 70°C for 5 h to form the membranes. For our experiment we used membranes with 1 mm x 1 mm in size, centered on 3 mm x 3 mm frames.

The chips were spincoated with HSQ (Hydrogen silsesquioxane) FOX16 resist with 3500 rpm resulting in an HSQ layer of 550 nm thickness. The exposure was done by 100 kV electron beam lithography using a Vistec EBPG 5000PlusES system. We used an aperture of 400 µm and a beam current of 1 mA. The exposure dose was varied from 7000 µC/cm² to 14500 µC/cm² in order to vary the HSQ duty cycle. The chip was developed in a 1:3 solution of AZ351-developer:$H_2O$ for 6 min, rinsed in water and dried by critical point drying in Leica CPD300 system in order to avoid structure collapse due to capillary forces. After cleaving the chip into single membranes, they were coated with Ir by plasma enhanced atomic layer deposition (ALD) in a Picosun R200 ALD tool. In our process, one cycle consisted of 3 pulses with 0.8 s of Ir(acac)3 as precursor and one pulse with 3 s of oxygen plasma (ICP power 2000 W, gas flow 50 sccm) as reactant [22], the temperature was set to 370°C during deposition.

## Experimental Setup

During fabrication of FZPs, means for quality control are essential. A very common method is inspection in a scanning electron microscope (SEM). However, lot of effort is required to control line width and duty cycle of FZPs over its full radius in order to identify the best fabrication parameters. Additionally, SEM only provides images of the surface, and cannot reveal the line widths and duty cycles for template structures with non-vertical side walls in the depth. Therefore structural changes from top to bottom having a significant effect on the efficiency of a FZP cannot be identified. Eventually, an efficiency measurement at a synchrotron provides the most reliable characterization. Unfortunately, access to synchrotron beam time is very sparse, and no option for a frequent, systematic investigation of zone plate performance.

As alternative we choose an X-ray tube for our measurements. Using lab sources for x-ray optics characterization has already been shown in several papers using a micro focus source [23] or a liquid-jet laser-plasma source [24]. In our experiment we used a conventional SEIFERT tube used under 6° in point focus geometry providing a spot size of approximately 0.4 mm x 0.8 mm.

We mounted a membrane chip with 16 FZPs in the beam of the tube, and placed a detector system with a YAG scintillator in the image plane of the FZPs (see Figure 6). The YAG crystal was imaged by a reverse mounted RICOH TV lens 12 mm via a 45° mirror on the CCD of a PCO Edge gold camera with a size of 2048 x 2048 pixels, each having a size of 6.5 µm x 6.5 µm. The distances were set such that a demagnified image of the source was focused on the detector. Using an x-ray tube with copper anode at 30 kV and 20 mA, filtered by with 12 µm thick nickel filters and 1 m air path, we obtained an effective spectrum absorbed by the 20 µm thick Yttrium-aluminum garnet (YAG) crystal that was centered around 8 keV. Such a system is suitable to compare the diffraction efficiency of FZPs in a relative manner, without providing absolute efficiency values.

All 16 zone plates on the membrane had the very same diameter of 150 µm, and an effective outermost half pitch of the Ir structures of 30 nm, but were exposed with a different electron beam exposure dose. As the exposure dose has a direct influence on the resulting line width of the HSQ template this provides an easy way to vary the HSQ duty cycle. The duty cycle of Ir was varied by the number of ALD cycles. To avoid any influence of local defects, the very same chip was repeatedly coated with around 2 nm Ir by ALD. After each deposition run, the coating thickness was inspected by SEM and the resulting focal spot brightness measured by the CCD using the x-ray tube.

## Results

At the X-ray tube the whole membrane including all 16 FZPs was imaged at once, as shown in Figure 6 above. The exposure time was 1 min. After dark field and flat field correction, the number of counts in the focal spot was calculated by summing up over a circle with 25 pixel diameter in the center of each zone plate projection. All the different values for the 16 FZPs, which differ in HSQ duty cycle, and all images from 27 coating steps, which differ in Ir duty cycle are drawn in Figure 7, representing a total number of 432 different zone geometries. For all the measurements the number of ALD cycles and the electron beam exposure dose was precisely known. These values were correlated to the line width measured with SEM. The duty cycles were calculated accordingly.

The recorded data show a clear maximum when varying the deposited Ir thickness. The position and height of this maximum depends on the HSQ duty cycle. For small HSQ duty cycles the maximum is reached later as for big HSQ duty cycles. The maximum number of counts was measured for a FZP with an HSQ duty cycle of 0.22 and an Ir duty cycle of 0.32, in excellent agreement with the theoretical predictions. The small deviation most likely results from SEM measurement uncertainties, but may also be a result of not visible structural changes in the bottom part of the FZP.

As we could only measure relative diffraction efficiency, a reference was needed. We used a FZP with identical diameter, zone width, and structure height, which had been characterized by an efficiency measurement using synchrotron radiation. At 8 keV the first order diffraction efficiency was 5.5 % [13]. The result of the reference measurement of this FZP at the X-ray tube is indicated by the black line. It is evident, that line-doubled zone plates made with the optimum duty cycles can provide substantially higher efficiency values.

# Conclusion

This systematic study of line-doubled FZPs has shown that an X-ray tube is a useful tool to quickly measure the diffraction efficiency of line-doubled Fresnel zone plates in a relative manner. Combining this type of measurement with an iterative variation of the fabrication parameters allowed us to identify the optimum fabrication parameters. This enables us to significantly improve the overall efficiency of line-doubled FZPs. With the X-ray tube we have an easy option for a non-destructive quality control without a synchrotron.

# Acknowledgment

The authors would like to thank C. Detlefs and T. Roth for their assistance at the ID06 beamline of the European Synchrotron Radiation Facility. This work received funding from the EU-H2020 Research and Innovation Programme under grant agreement No 654360 NFFA-Europe.

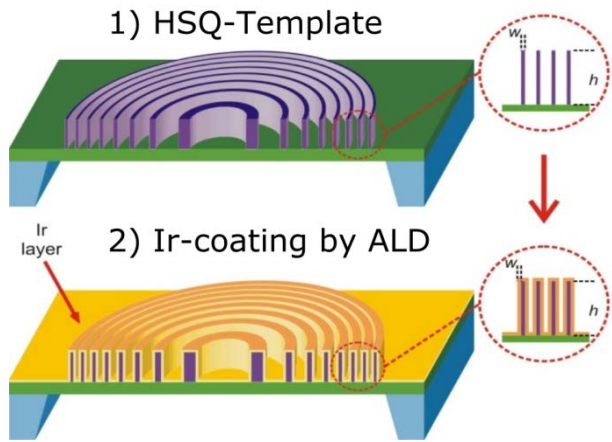

Figure 1: Schematic of the fabrication process of line-doubled zone plates (blue: Si-frame, green: Si$_3$N$_4$, purple: HSQ, yellow: Ir) [13].

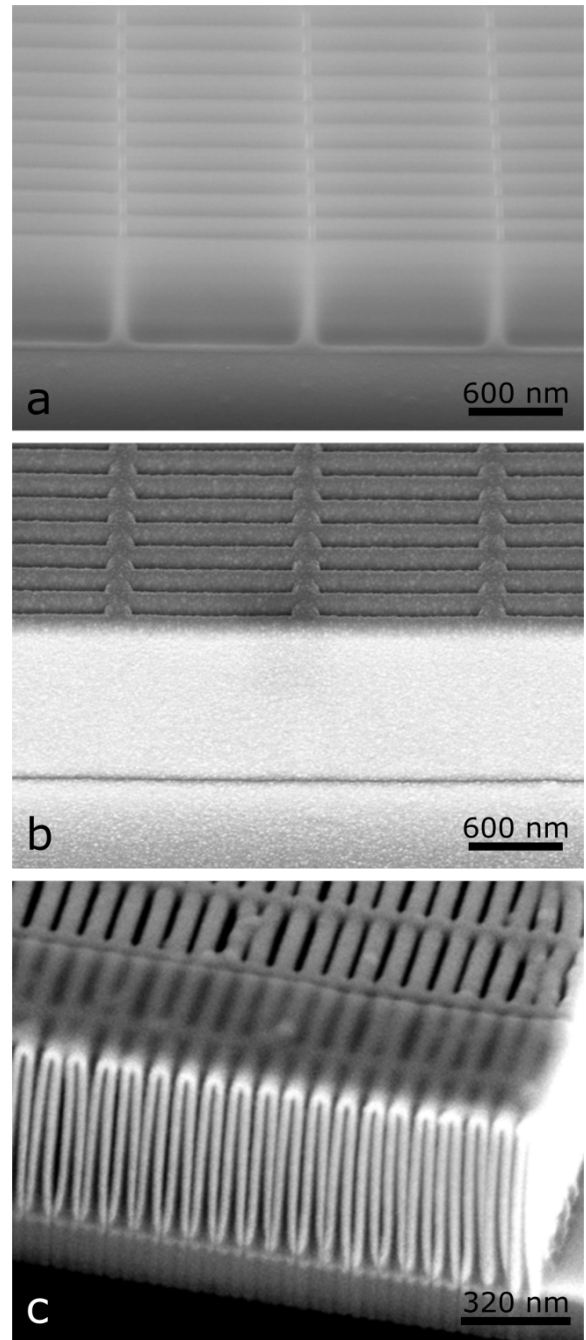

Figure 2: SEM images of a line-doubled zone plate with 30 nm half Ir pitch and 60 nm half HSQ pitch: a) HSQ-template, b) after Ir-coating. And c) FIB cross section of the Ir coated HSQ template with 20 nm half Ir pitch.

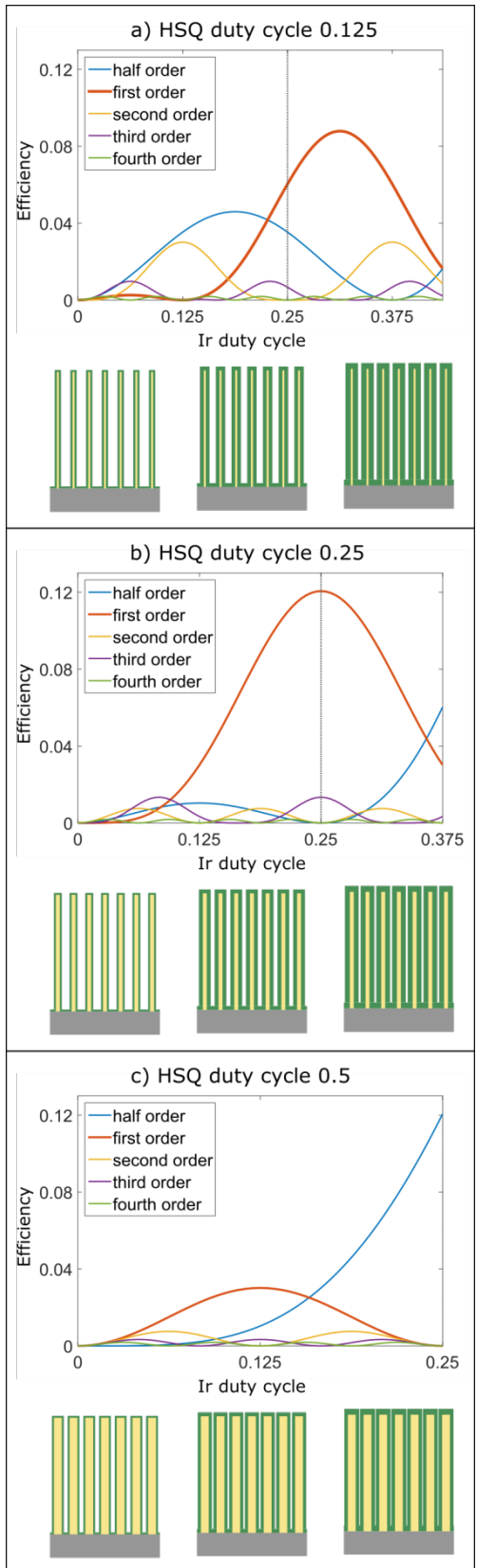

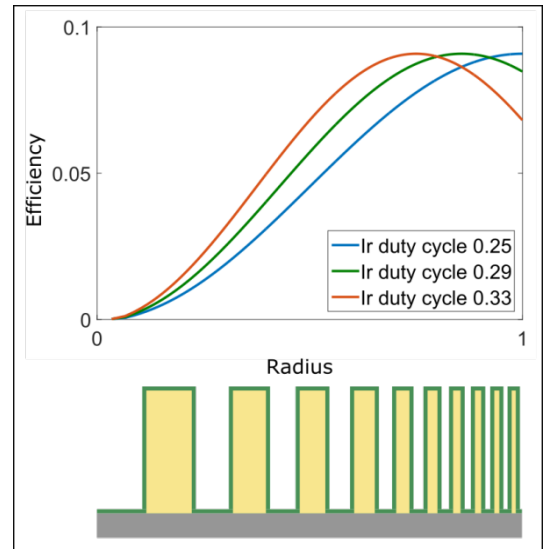

Figure 4: Dependence of the diffraction efficiency of a line-doubled zone plate over its radius.

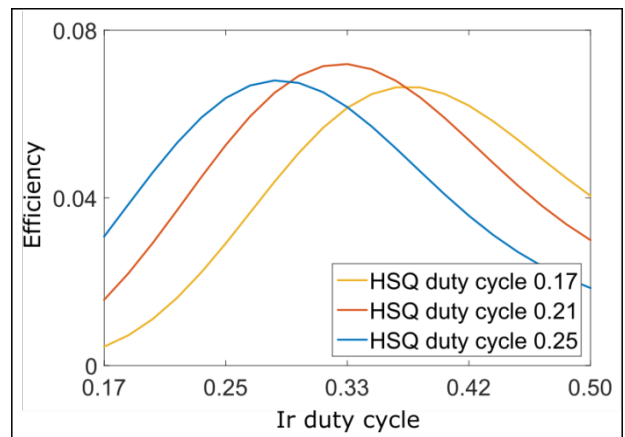

Figure 5: Efficiency of a line-doubled zone plate versus duty cycle of HSQ and Ir.

Figure 3: Diffraction efficiency of line-doubled gratings as a function of HSQ duty cycle and Ir duty cycle.

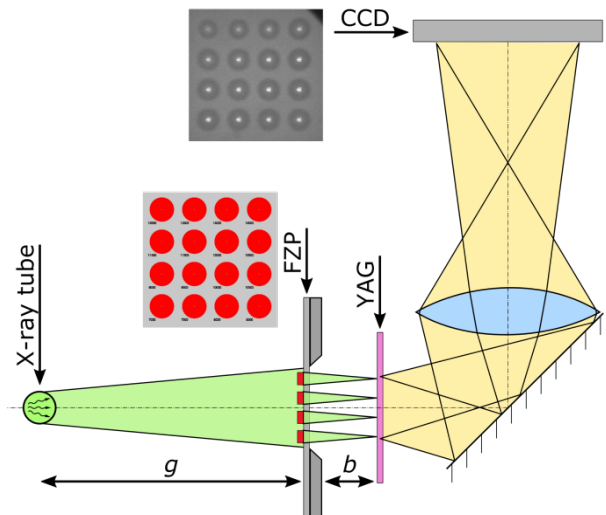

Figure 6: Schematic of the experimental setup for efficiency measurements on an X-ray tube. The distance between source and FZP *g* is 1 m and the distance between FZP and YAG *b* is 28 mm.

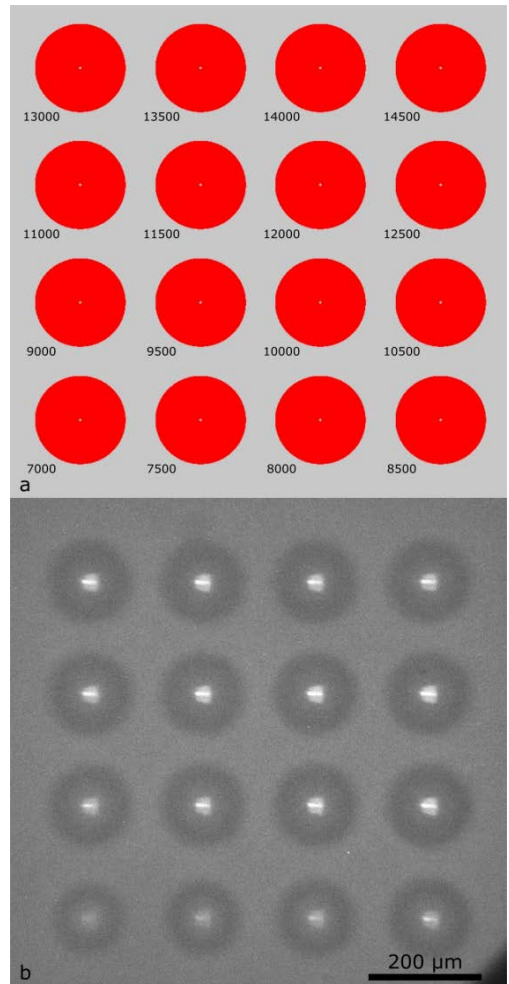

Figure 7: a) Layout of the membrane with 16 FZP and their doses in µC/cm² and b) x-ray image in the focal plane of the chip coated with 1400 cycles Ir in the ALD. For both, the layout and the x-ray image the zone plate diameter is 150 µm and the pitch of the zone plate array is 200 µm.

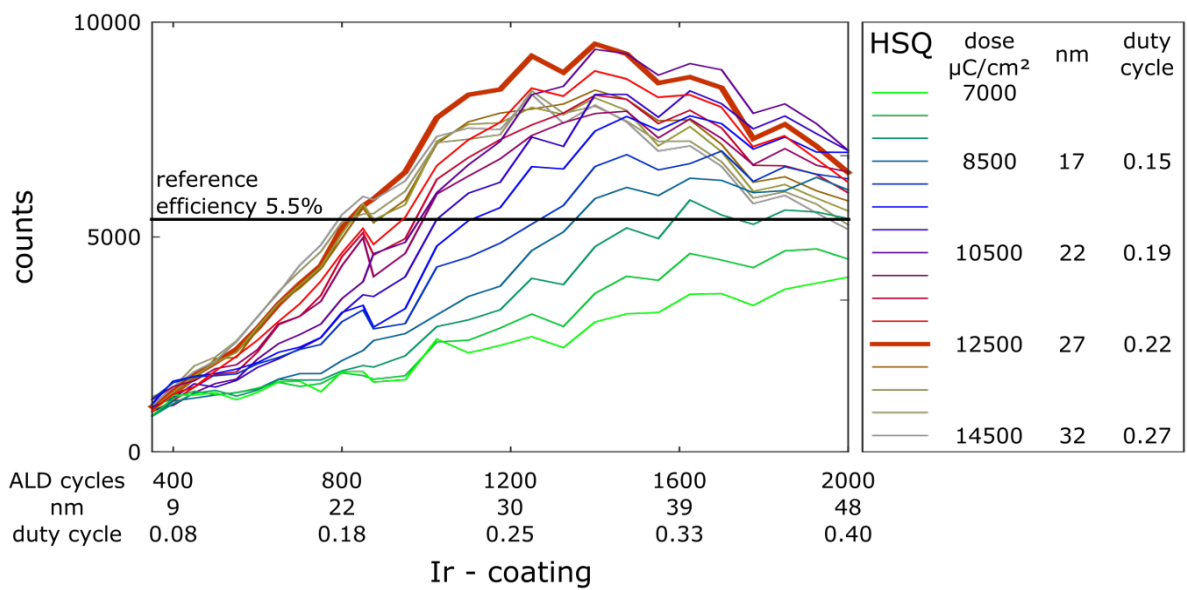

Figure 8: Measured counts in focus dependent on HSQ and Ir duty cycles. Both duty cycles are calculated from the measured line width at SEM and correlated to the electron beam exposure dose or the number of ALD cycles,

respectively. The black line indicates the counts of a reference zone plate with the same design, which was characterized with synchrotron radiation, showing an efficiency of 5.5% @ 8 keV.